\documentclass[12pt]{article}
\usepackage{pdproc,epsfig} 
\begin{document}

\def\lesssim{\mathrel{\mathop
  {\hbox{\lower0.5ex\hbox{$\sim$}\kern-0.8em\lower-0.7ex\hbox{$<$}}}}}
\def\greatsim{\mathrel{\mathop
  {\hbox{\lower0.5ex\hbox{$\sim$}\kern-0.8em\lower-0.7ex\hbox{$>$}}}}}
\def\nn{\nonumber}
\def\ee{\end{equation}}
\def\be{\begin{equation}}
\def\eea{\end{eqnarray}}
\def\bea{\begin{eqnarray}}
\def\eeas{\end{eqnarray*}}
\def\beas{\begin{eqnarray*}}
\renewcommand{\thefootnote}{\fnsymbol{footnote}} 

\def\lapprox{\mathrel{\mathop
  {\hbox{\lower0.5ex\hbox{$\sim$}\kern-0.8em\lower-0.7ex\hbox{$<$}}}}}
\def\gapprox{\mathrel{\mathop
  {\hbox{\lower0.5ex\hbox{$\sim$}\kern-0.8em\lower-0.7ex\hbox{$>$}}}}}

\renewcommand{\thefootnote}{\alph{footnote}}

\title{A few comments after the charged current measurement at the Sudbury Neutrino Observatory}

\author{G. Fiorentini, F.L. Villante and B. Ricci}

\address{Dipartimento di Fisica, Universit\`a di Ferrara and INFN-Sezione di
Ferrara, I-44100 Ferrara\\
 {\rm E-mail: fiorentini@fe.infn.it, villante@fe.infn.it, ricci@fe.infn.it}}

\abstract{The comparison of the SNO charged current result with the 
 solar neutrino signal measured by Super-Kamiokande
 has provided, for the first time,
the evidence of a non electron flavour active neutrino component
in the solar flux. We remark here that this evidence
can be obtained in a model independent way, i.e. without any assumpion
about solar models, about the energy dependence of the neutrino oscillation 
probability and about the presence of sterile neutrinos.
Furthermore, from the $^8$B neutrino flux obtained by combining
SNO and Super-Kamiokande, one can determine
the central solar temperature, 
$T=1.57\cdot(1\pm 1\%)\cdot 10^{7}$~K, 
and provide an estimate of 
the beryllium neutrino flux, 
$\Phi_{\rm Be}=4.9\cdot(1\pm 11\%) \cdot 10^9 \;{\rm cm}^{-2} {\rm s}^{-1}$.}

\normalsize\baselineskip=15pt

\section{Introduction}

All solar neutrino experiments performed over the past thirty years and 
 probing different portions of the solar neutrino spectrum
 \cite{Hom,Gal,Sage,Kam,SK}
have reported a deficit with respect to the Standard Solar Model (SSM),
 see Table \ref{Tab1}.
 Furthermore, by combining the various
experimental results, it has been possible to obtain indications in favour
of non standard neutrinos even independently of SSM, see e.g. \cite{Report}.
All this evidence, however
is not a proof, and the focus is now on the search of real footprints
of neutrino oscillations. 

In this respect, SNO \cite{SNO} has recently produced an important result,
i.e., when combined with Super-Kamiokande data, SNO provides
evidence of a non electron flavour active neutrino component
in the solar flux. Briefly, as it was suggested in ref. \cite{tio3},
 one can use the charged currents (CC) at SNO to tell neutral currents
at Super-Kamiokande (SK).

 The suggestion of ref. \cite{tio3}  was based on the following points:
\begin{itemize}
\item
If $\nu_e$ oscillate into active neutrinos of different flavour
-- say $\nu_{\mu}$ -- these too contribute to the Super-Kamiokande
solar neutrino signal $S$;
\item
The contribution $S_{\mu}$ of muon neutrinos to SK signal
 is expected to be much larger than the experimental error
$\Delta S$;
\item
By a suitable choice of the energy threshold, the CC signal of SNO can
 be used to determine the $\nu_{e}$ component $S_{e}$ of the SK signal;
\item
The accuracy of SNO should be sufficient for extracting the $\nu_{\mu}$
contribution to the SK signal, from $S_{\mu}=S-S_{e}$.

\end{itemize}

All this means that  by combining the results of SK with CC data from
SNO, one can have a signature of oscillations among active neutrinos. 

{\em An important point is that this
signature is  independent of assumptions about the solar model and
the neutrino spectrum.}

In the first part of this paper, following \cite{tio3}
 we shall present a general
derivation of this result, which has been used in \cite{SNO,elio,giunti}
in order to perform a model independent analysis of 
experimental data
\footnote{For a different approach to the analysis of SK and SNO combined data see also
\cite{venya}}. The method, outlined here for the energy
integrated rates, has been extended in ref. \cite{Fogli}
for the comparison of the SK and SNO energy spectra.

As also suggested in \cite{tio3},
by combining the CC results of SNO and SK 
it is  possible to determine experimentally the total active
flux of $^{8}$B neutrinos produced in the sun.
This too has been succesfully accomplished  by SNO \cite{SNO},
 which has thus 
provided a refined test of the SSM and has opened
the possibility of using neutrinos as probes of the sun. 

In the final part of this paper we shall comment 
on the accuracy of neutrinos as thermometer of the solar interior.
Furthermore, we shall provide a determination of the $^7$Be neutrino
flux, which is essentially independent of the solar model.


\begin{table}[h]
\caption{Experimental results 
of solar neutrinos experiments, compared with theoretical predictions.}
\vspace{5 mm}
\small
\begin{tabular}{||l|l|c|c||}
\hline\hline
Reaction&
Experiment & SSM prediction  \cite{BP01} & Experimental result\\\hline
&&&\\
&GALLEX+GNO&                          & $74.1^{+6.7}_{-6.8}$ SNU  \\
$\nu_e+^{71}$Ga$\rightarrow ^{71}$Ge$+e^-$  &&$128^{+9}_{-7}$ SNU       & \\
&SAGE      &                          & $67.2 ^{+ 8.0}_{-7.6}$ SNU \\
&          &&\\
\hline
&&&\\
$\nu_e+^{37}$Cl$\rightarrow ^{37}$Ar$+e^-$ & 
Homestake & $7.6^{+1.3}_{-1.1}$ SNU  & $2.56\pm0.22$ SNU  \\
&&&\\
\hline
&&&\\
 & Kamiokande&                 & $(2.80 \pm 0.38) \cdot 10^6$ cm$^{-2}$s$^{-1}$\\
&&&\\
$\nu+e^- \rightarrow \nu+e^-$&
Super-Kamiokande & 
$5.05 (1^{+0.20}_{-0.16}) \cdot 10^6$cm$^{-2}$s$^{-1}$  &
$(2.32 \pm 0.08) \cdot 10^6$ cm$^{-2}$s$^{-1}$\\
&&&\\
& SNO    &                    & $(2.39 \pm 0.37) \cdot 10^6$ cm$^{-2}$s$^{-1}$  \\
&&&\\
\hline
$\nu_e+d \rightarrow p+p+e^-$ & 
SNO    
&  $5.05 (1^{+0.2}_{-0.16}) \cdot 10^6$cm$^{-2}$s$^{-1}$                  
& $(1.75 \pm 0.14) \cdot 10^6$ cm$^{-2}$s$^{-1}$ \\
&&&\\
\hline  \hline
\end{tabular}
\label{Tab1}
\end{table}


\section{The SNO-Super-Kamiokande connection}  

The goal of this section is to show that one can extract 
from the Super-Kamiokande data
the contribution
of active neutrinos different from $\nu_e$ 
by using the CC  result of SNO, {\em without assumptions about the solar
model and the neutrino spectrum.}

In order that the argument can be presented in a very simple way, 
first  we assume that the solar neutrino spectrum is not distorted.
i.e. the survival (transition) probability $P_{ee}$ ($P_{e\mu}$)
 is energy independent. Next we shall prove that
our results hold generally, i.e. independently of the functional form of
$P_{ee}(E_{\nu})$ and $P_{e\mu}(E_{\nu})$.

\subsection{The case of energy independent oscillations}

Super-Kamiokande detects electrons from the reaction
\begin{equation}
\nu_{l}+e^{-}\rightarrow\nu_{l}+e^{-} \quad \quad
l=e,\;\mu,\;\tau
\label{a1}
\end{equation}
with kinetic
\footnote{Super-Kamiokande generally quotes the ``visible energy''
$E_{vis}$ of the detected electron, which corresponds to the
total electron energy ($E_{vis}=T+m_{e}$).}  
energy $T\ge T_{\rm SK}=4.5\;{\rm MeV}$.

After  1258 days of operations, the  signal
is $S=(0.459\pm0.016)\;S^{\rm (SSM)}$ \cite{SK}, where $S^{\rm(SSM)}$ is
the prediction of the Standard Solar Model of ref \cite{BP01},
corresponding to a  $^{8}$B neutrino flux 
$\Phi_{\rm B}^{\rm (SSM)}=5.05\cdot 10^{6}$ cm$^{-2}$ s$^{-1}$. Statistical
(1$\sigma$) and systematical errors, here and in the following,
are added in quadrature.

If the neutrino spectrum is undistorted, the contribution of $\nu_{e}$
and $\nu_{\mu}$ to the SK signal
(here and in the following we use the index $\mu$
for any active neutrino species different from $\nu_{e}$),
\begin{equation}
S=S_{e}+S_{\mu}
\label{a2}
\end{equation}
are simply given by 
\begin{equation}
S_{e}= \Phi_{\rm B} \; \overline{\sigma}_e\;P_{ee}
\quad \quad  ;  \quad \quad
S_{\mu}=\Phi_{\rm B} \; \overline{\sigma}_\mu\;P_{e\mu}
\label{a3} 
\end{equation}
where $\Phi_{\rm B}$ is the flux of boron neutrinos produced in the sun,
 $\overline{\sigma}_{e}$ and 
$\overline{\sigma}_{\mu}$ are the effective cross sections for
$\nu_e$ and $\nu_{\mu}$ detection (see the next subsection for a precise
definition) and
all signal rates, here and in the following, 
are normalized to unit target. 
The ratio,
\begin{equation}
\beta=\frac{\overline{\sigma}_\mu}{\overline{\sigma}_e}\;,
\label{a4}
\end{equation}
in the energy range of interest to us is $\beta\simeq0.152$.

Clearly SK determines only the sum of $S_{e}$ and $S_{\mu}$
and cannot discriminate between the two contributions.

SNO has recently reported results on the rate 
of solar neutrino events from the charged current reaction:
\begin{equation}
\nu_{e}+d\rightarrow p+p+e^{-} ~.
\label{a5}
\end{equation}
The observed rate is $C=(0.347\pm0.029)\;C^{(SSM)}$ \cite{SNO}where $C^{(SSM)}$
is the SSM-predicted signal.

If the neutrino spectrum is undeformed, the $\nu_{e}$ flux 
at earth ($\Phi_e=\Phi_{\rm B} P_{ee}$)
can be determined from:
\begin{equation} 
C = \Phi_{\rm B}\;\overline{\sigma}_{\rm CC}\;P_{ee}  ~.
\label{a6}
\end{equation}
One can thus extract  the muon signal in SK by combining the SK and SNO
experimental results. From the previous equations one has:
\begin{equation} 
S_{\mu}=
S-\frac{\overline{\sigma}_{e}}{\overline{\sigma}_{\rm CC}}\;C
\label{a7}
\end{equation}
Moreover, under the hypothesis of oscillation only among active neutrinos,
one can also get the total flux of neutrinos
produced in the sun.
The foregoing equations give in fact 
\begin{equation} 
\Phi_{\rm B} =
\left(\frac{\beta-1}{\beta}\;\frac{C}{\overline{\sigma}_{\rm CC}}
+\frac{1}{\beta}\;\frac{S}{\overline{\sigma}_{e}}\right)
\label{a8}
\end{equation}
{\em We remark that $S_{\mu}$ and $\Phi_{\rm B}$ are determined, by means of 
(\ref{a7}) and (\ref{a8}), only in terms of experimental data}, 
independently of solar models, assuming however that the survival/transition
probability is energy independent.

By using  the available data from SK, the SSM  prediction
for $\Phi_{\rm B}$ and a conservative ($\sim 10\%$) estimate
of the uncertainty of  $\overline{\sigma}_{\rm CC}$, 
in ref. \cite{tio3} it was concluded that: 
\begin{itemize}
\item
A muon signal $S_{\mu}=0.10\,S_{\rm BP}$,
should be evident at about 2$\sigma$ by combining the SK results
with the SNO data taken during one year.
\item
The total $^{8}$B flux  should  be measured with an accuracy of about
20\% during the same time.
\end{itemize}

The beautiful measurement of SNO \cite{SNO} has confirmed these predictions.
In fact the comparison of the CC signal of SNO to the SK result
has shown \cite{SNO} a $3.3\sigma$ difference,
providing thus a significant evidence that there is a non-electron flavour
active neutrino component  in the solar flux
\footnote{The $3.3\sigma$ significance instead of $2\sigma$ predicted
in \cite{tio3} is due mainly to a more recent ($\sim 3\%$) estimate of 
the uncertainty of  $\overline{\sigma}_{\rm CC}$, see \cite{SNO} and references therein.}.

The total flux of active $^8$B neutrinos has been determined
by the SNO collaboration as:
\begin{equation}
\label{fibsno}
\Phi_{\rm B}=(5.44 \pm 0.99) \cdot 10^6 \,{\rm cm}^{-2}\,{\rm s}^{-1} \quad , 
\end{equation}
in close agreement
with the predictions of solar models, see Table \ref{Tablessm}.

\subsection{The general case}

In the general case the transition/survival probability depends on the neutrino
energy and  eqs. (\ref{a3})
and (\ref{a6}) do not hold. Nevertheless  due to the fact that 
{\it the SK and SNO response functions can be equalized the final relations
(\ref{a7}) and (\ref{a8}) still hold.}

Let us explain this statement in some detail. In the general case,
eqs. (\ref{a3}) and (\ref{a6}) are replaced by:
\begin{eqnarray}
\label{a10} 
S_{e}   &=& \Phi_{\rm B} \; \overline{\sigma}_e\;
\langle P_{ee} \rangle_{\rm SK} ~,\\
\label{a11}
S_{\mu} &=&  \Phi_{\rm B} \;\overline{\sigma}_{\mu}\;
\langle P_{e\mu} \rangle_{\rm SK} ~,\\
C &=& \Phi_{\rm B} \;\overline{\sigma}_{\rm CC}\;
\langle P_{ee} \rangle_{\rm SNO}  ~,
\label{a12}
\end{eqnarray}
where the energy averaged oscillation probability are:
\begin{eqnarray} 
\label{a13}
\langle P_{ee} \rangle_{\rm SK} &=& 
\int_{0}^{\infty} dE_{\nu} \; P_{ee}(E_{\nu})\; \rho_{e}(E_{\nu},T_{\rm SK}) ~,\\
\label{a14}
\langle P_{e\mu} \rangle_{\rm SK} &=& 
\int_{0}^{\infty} dE_{\nu} \; P_{e\mu}(E_{\nu})\; \rho_{\mu}(E_{\nu},T_{\rm SK}) ~,\\
\langle P_{ee} \rangle_{\rm SNO} &=& 
\int_{0}^{\infty} dE_{\nu} \; P_{ee}(E_{\nu}) \;\rho_{\rm CC}(E_{\nu},T_{\rm SNO}) ~.
\label{a15}
\end{eqnarray}


\begin{figure}
\begin{center}
\epsfig{file=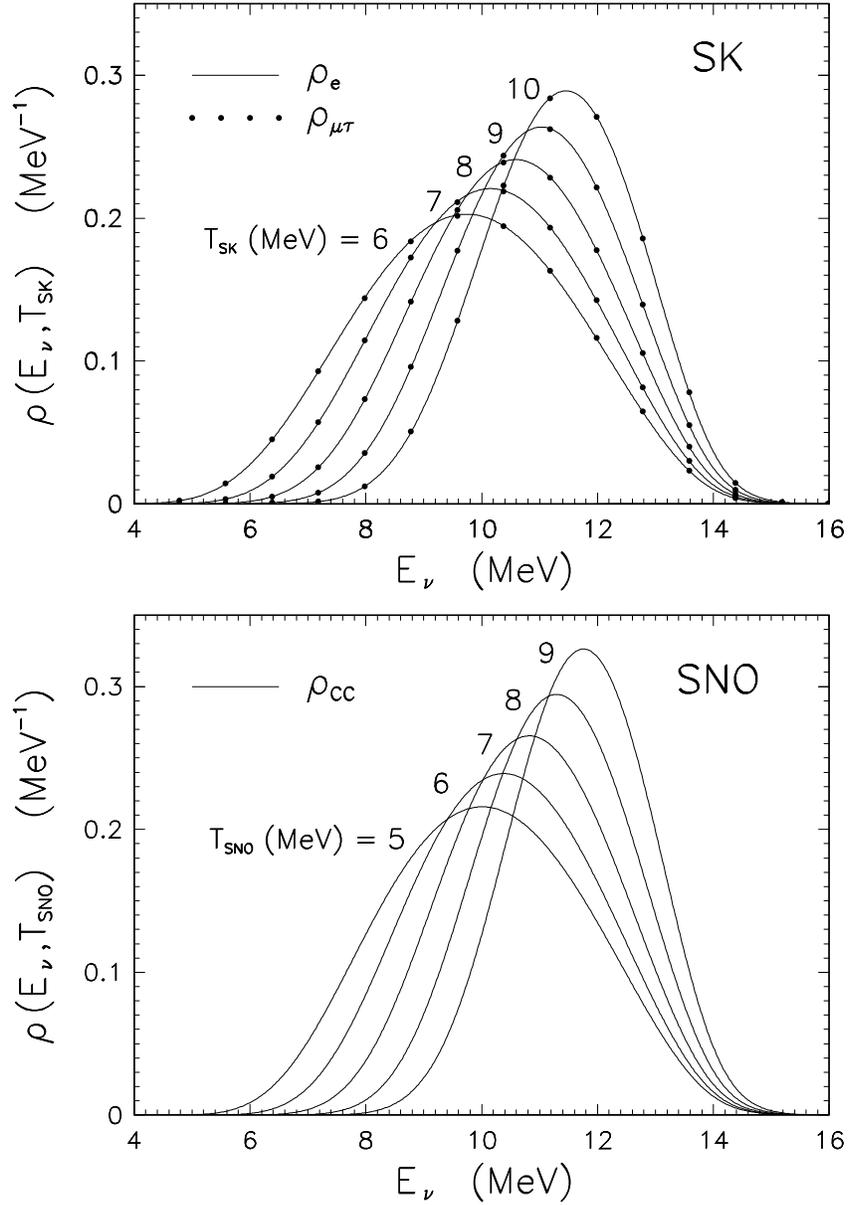,height=17cm}
\end{center}
\caption{The normalized response functions of SK and SNO to
$^{8}B$ neutrinos, for representative values of detector
thresholds. See text for details.}
\label{f20}
\end{figure}

The response functions  ($\rho_e$, $\rho_\mu$ and $\rho_{\rm CC}$)
summarize completely the SK and SNO
detector properties. They are defined in terms of the 
standard $^8$B energy spectrum,
$\lambda_{\rm B}(E_{\nu})$ \cite{Bspe},
of the energy thresholds ($T_{\rm SK}$ and $T_{\rm SNO}$)
in the two experiments,
of the differential cross sections for the detection processes
($d\sigma_e/dT'$ and $d\sigma_{\mu}/dT'$
for elastic scattering \cite{EScs} and $d\sigma_{\rm CC}/dT'$ for CC absorption \cite{CCcs})  
and of the detector resolution functions ($r_{\rm SK}(T,T')$ \cite{SKre}
and $r_{\rm SNO}(T,T')$ 
\footnote{In this work we use the SNO resolution function described in
\cite{tio3} and references therein. See \cite{SNO} for a more 
recent determination.}:
\begin{eqnarray}
\rho_{e} &=& \frac{\displaystyle
\lambda_{\rm B}(E_\nu)\int_{T_{\rm SK}}^{\infty}dT
\int_0^{\infty}dT'\,
\frac{d\sigma_{e}(E_{\nu},\,T')}{dT'}\,r_{\rm SK}(T,\,T')}
{\overline{\sigma}_{e}(T_{\rm SK})}\ ,
\label{rhoe}\\[4mm]
\rho_{\mu} &=& \frac{\displaystyle
\lambda_{\rm B}(E_\nu)\int_{T_{\rm SK}}^{\infty}dT
\int_0^{\infty}dT'\,
\frac{d\sigma_{\mu}(E_{\nu},\,T')}{dT'}\,r_{\rm SK}(T,\,T')}
{\overline{\sigma}_{\mu}(T_{\rm SK})}\ ,
\label{rhoa}\\[4mm]
\rho_{\rm CC} &=& \frac{\displaystyle
\lambda_{\rm B}(E_\nu)\int_{T_{\rm SNO}}^{\infty}dT
\int_0^{\infty}dT'\,
\frac{d\sigma_{\rm CC}(E_\nu,\,T')}{dT'}\,r_{\rm SNO}(T,\,T')}
{\overline{\sigma}_{\rm CC}(T_{\rm SNO})}\ 
\label{rhoc}
\end{eqnarray}
The denominators $\overline{\sigma}_{e}$ ($\overline{\sigma}_{\mu}$) and 
$\overline{\sigma}_{\rm CC}$ represent the total cross sections for 
electron (muon) neutrino detection in SK and SNO respectively,
as obtained by integrating over $E_\nu$ the corresponding numerators in
Eqs.~(\ref{rhoe})--(\ref{rhoc}). The response
functions are normalized to unity.

Figure~1 shows the response functions of SK and SNO 
as a function of the neutrino energy, for representative values
of the detector thresholds. 
One sees that 
$\rho_e$ and $\rho_{\mu}$ are almost coincident.
This is not surprising  
since the cross  sections for $\nu_e\,e$ and $\nu_{\mu}e$ 
scattering have a similar shape in the range probed by SK, 
up to an overall factor \cite{Hi87}. For any
practical purpose, one can assume that
\begin{equation}
\rho_e(E_\nu,T_{\rm SK})=\rho_{\mu}(E_\nu,T_{\rm SK})
\label{a19}
\end{equation}
to a very good approximation. 

It is intriguing to notice that the SK and SNO response functions in Fig.~1
also look very similar, provided that the SNO threshold is chosen
 1-2 MeV below the SK threshold, despite the fact that the differential 
cross sections are very different.
This feature can be understood, qualitatively, by means of the following 
considerations:
\\
{\it i)} The high energy behaviour of the response functions is essentially
determined by the shape of the $^8$B neutrino spectrum, and is almost
unsensitive to the cross section. In particular,
in the limit of infinite resolution, the fact that the $^8$B
neutrino spectrum has an end-point at $E_{\nu} \simeq 14$~MeV gives:
\begin{eqnarray}
\rho_{e,\mu}(E_\nu,T_{\rm SK}) &=& 0 \quad {\rm for}\quad  E_\nu \geq 14 ~{\rm MeV} 
\nonumber
\\
\label{eqless11}
\rho_{\rm CC}(E_\nu,T_{\rm SNO}) &=& 0 \quad {\rm for} \quad E_\nu \geq 14 ~{\rm MeV}
\nonumber
\label{eqless22}
\end{eqnarray}
{\it ii)} The low energy behaviour depends on the chosen threshold; i.e.,
in the limit of infinite resolution:
\begin{eqnarray}
\rho_{e,\mu}(E_\nu,T_{\rm SK}) &=& 0 \quad {\rm for}\quad  E_\nu \leq T_{\rm SK}+m_e/2 
\nonumber
\label{eqless1}\\
\rho_{\rm CC}(E_\nu,T_{\rm SNO}) &=& 0 \quad {\rm for} \quad E_\nu \leq T_{\rm SNO}+Q
\nonumber
\label{eqless2}
\end{eqnarray}
where $Q=2m_p+m_e-m_d=1.442$ MeV.\\
{\it iii)} This implies that, by taking $T_{\rm SK}-T_{\rm SNO}\simeq Q -m_e/2 \simeq 1.2$ MeV,
the response functions of SK and SNO are non vanishing in the same neutrino 
energy window. Since both are normalized to unit area, thei heights have thus to be
similar.

The presence of a finite resolution makes the picture more
complicated. Numerically, the relative position of the two
thresholds has been optimized so as to minimize the difference between 
the response functions.

\begin{figure}
\begin{center}
\epsfig{file=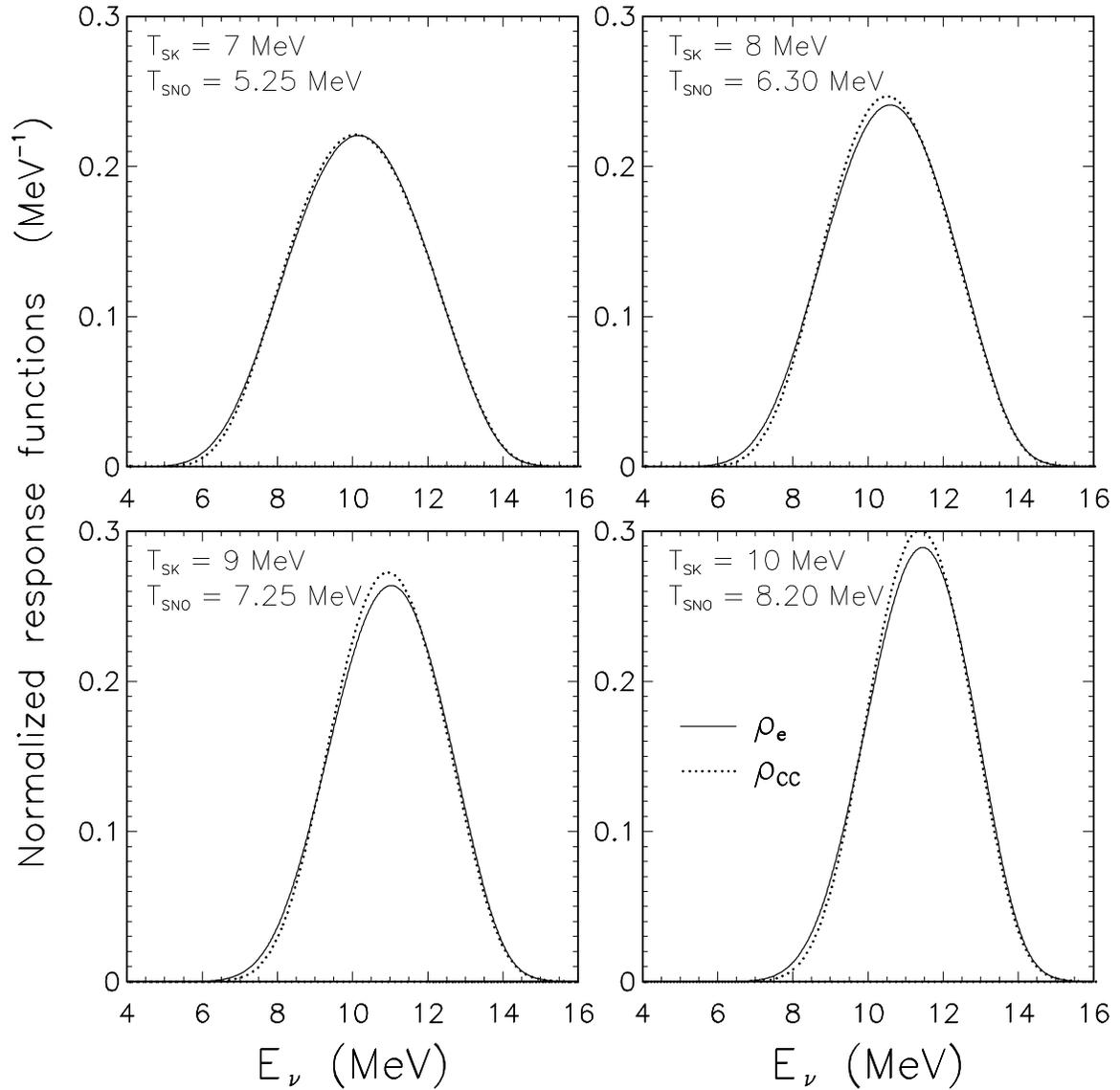,height=16cm}
\end{center}
\caption{Examples of the approximate equality of the SK and SNO
response functions for selected values of the detector thresholds
$T_{\rm SK}$ and $T_{\rm SNO}$. Such values obey approximately the
relation $T_{\rm SNO}=0.995\,T_{\rm SK}-1.71 ({\rm MeV})$.}
\label{f30}
\end{figure}

The results are shown in Fig.~2 for some representative values of the
thresholds. One sees that the 
two response functions $\rho_e$ and $\rho_{\rm CC}$ 
can be equalized to a good approximation. 
Details can be found in ref.\cite{tio3}.
Briefly , in the calculations of the electron rates $S$ and
$C$ one can assume that {\em the SK and SNO response functions are
equal}
\begin{equation}
\rho_{\mu}(E_\nu,T_{\rm SK}) = \rho_e(E_\nu,T_{\rm SK}) = \rho_{\rm CC}(E_\nu,T_{\rm SNO})~,
\label{a23}
\end{equation}
{\em within an accuracy of a few percent or less, provided that
the SK and SNO thresholds are chosen according to the empirical relation}.
\begin{equation}
T_{\rm SNO} = 0.995\, T_{\rm SK} - 1.71 {\rm\ \ (MeV)}\ .
\label{a24}
\end{equation}

 The equality~(\ref{a23}) allows 
a generalization of the results obtained in the previous section.
From equations (\ref{a14}) and (\ref{a15}), 
{\it independently of the functional form of $P_{ee}(E_{\nu})$}
one has:
\begin{equation}
\langle P_{ee} \rangle_{\rm SK} = \langle P_{ee}\rangle_{\rm SNO}~.
\label{a25}
\end{equation}
This shows  that the $\nu_{e}$ contribution to SK signal can  be determined
from the CC rate at SNO:
\begin{equation}
\frac{S_{e}(T_{\rm SK})}{\overline{\sigma}_{e}(T_{\rm SK})} =
\frac{C(T_{\rm SNO})}{\overline{\sigma}_{\rm CC}(T_{\rm SNO})}~.
\label{a26}
\end{equation}
As a consequence, the muon contribution can
be extracted by combining data of the two experiments:
\begin{equation}
S_{\mu}(T_{\rm SK})=S(T_{\rm SK})-\frac{\overline{\sigma}_e(T_{\rm SK})}
{\overline{\sigma}_{\rm CC}(T_{\rm SNO})}\;C(T_{\rm SNO}) ~.
\label{a27}
\end{equation}
Moreover, under the assumption of oscillations only among active neutrinos,
$P_{ee}(E_{\nu})+P_{e\mu}(E_{\nu})= 1$, one obtains from eqs.~(\ref{a13})and
(\ref{a14}):
\begin{equation}
\langle P_{ee} \rangle_{\rm SK} = 1 - \langle P_{e\mu}\rangle_{\rm SK}~.
\label{a20}
\end{equation}
This relation, together with eqs.(\ref{a10}), (\ref{a11}), (\ref{a3}) and (\ref{a26}) 
gives the following expression for the total flux $\Phi_{\rm B}$:
\begin{equation}
\Phi_{\rm B}=
\left[\frac{\beta(T_{\rm SK})-1}{\beta(T_{\rm SK})}\;
\frac{C(T_{\rm SNO})}{\overline{\sigma}_{\rm CC}(T_{\rm SNO})}
+\frac{1}{\beta(T_{\rm SK})}\;\frac{S(T_{\rm SK})}
{\overline{\sigma}_{e}(T_{\rm SK})}\right]
\label{a28}
\end{equation}

We conclude thus that the basic equations (\ref{a7}) and (\ref{a8})
can be recovered in the general case, provided only 
that the SK and SNO thresholds are chosen accordingly to the empirical
relation (\ref{a24}).

The model independent approach described above was used
in ref. \cite{SNO,elio,giunti} to determine the contribution
$S_{\mu}$ of non electron active neutrino flavour components to SK signal.
%
It was concluded that $S_{\mu} > 0$, with a significance 
greater than $3\sigma$.
We remark that {\em this conclusion is independent of any assumption about the solar 
models, about the energy dependence of oscillation probability,
and about the presence of sterile neutrinos}.

Finally, under the additional assumption of oscillations only among active
neutrinos, by using eq.~(\ref{a28}) it was found in ref.\cite{elio} that
\begin{equation}
\Phi_{\rm B} = 5.20\cdot(1^{+0.20}_{-0.16}) \cdot 10^6\;\;{\rm cm}^{-2}~{\rm s}^{-1}~. 
\label{flux}
\end{equation}

This determination is more accurate, in principle than that of ref.\cite{SNO},
eq. (\ref{fibsno}), since it has been obtained by matching the SK and SNO
response function.

\section{Neutrinos as probes of the solar interior}
\label{subsectherm}

The central temperature $T$ of the  sun is a nice example of a physical 
quantity which can be determined by means of solar neutrino detection, 
provided  that the relevant nuclear physics  is known.
SSM   calculations predict $T$ with an accuracy of 1\% or even 
better, see e.g. \cite{BPBC}. 
However, this is a theoretical prediction which, as any 
result in physics, demands observational evidence.

The fluxes  of $^8$B and $^7$Be neutrinos are given by, 
see \cite{Report}:
\begin{eqnarray}
\label{eq2a}
        \Phi_{\rm B} &=& c_{\rm B}  S_{17} \frac{S_{34}}{\sqrt{S_{33}}} T^{20} \\
\label{eq2b}
 \Phi_{\rm Be} &=& c_{\rm Be} \frac{S_{34}}{\sqrt{S_{33}}} T^{10} 
\end{eqnarray}
where $S_{ij}$ are the low energy astrophysical factors for nuclear reactions 
between nuclei with atomic mass numbers i and j.
These quantities have been subject of intense experimental study in the last
few years. The present knowledge is summarized in Table \ref{Tablesij}.
Each factor is known with an accuracy of 10\% or better.


\begin{table}[h]
\caption[cc]{Zero energy astrophysical $S$ factors according to:
a) the compilation of Adelberger et al 
\cite{Adelberger};
b) the compilation of NACRE collaboration
 \cite{Nacre};
c) recent measurements not included in a) and b) 
The $1\sigma$ errors are indicated.}
\vspace{5 mm}
\small
\begin{center}
\begin{tabular}{||l|c|c|c||}
\hline \hline
   & a) & b) & c)\\
\hline
$S_{17}$ [eVb]   &    $19^{+4}_{-2}$   &   $21\pm 2$   & $18.8\pm 1.7$ \cite{s17direct}\\
                 &                     &               &  $17.8^{+ 1.4}_{-1.2}$ \cite{s17indirect}\\
\hline
$S_{33}$ [MeVb]   &   $5.40\pm 0.40$   &   $5.40\pm0.40$   & $5.32\pm 0.40$ \cite{s33luna}\\
\hline
$S_{34}$ [KeVb]   &    $0.53\pm 0.05$   &   $0.54\pm 0.09$   & \\
\hline \hline
\end{tabular}
\end{center}
\label{Tablesij}
\end{table}

The constants 
$c_{\rm B}$ and $c_{\rm Be}$ are well 
determined. With convenient units, the above equations become:
\begin{eqnarray}
\label{eq2ua}
        \Phi_{\rm B}({\rm cm}^{-2}{\rm s}^{-1}) &=&  1.443\cdot 10^{-18}\;
                   \left ( \frac{ S_{17}}{\rm{eVb}} \right ) \;
                   \left ( \frac{ S_{34}}{\rm{KeVb}} \right ) \;
                   \left ( \frac{ S_{33}}{\rm{MeVb}} \right )^{-1/2}  \; T_6^{20} \\
\label{eq2ub}
 \Phi_{\rm Be} ({\rm cm}^{-2}{\rm s}^{-1}) &=& 2.328\cdot 10^{-2} \;
                  \left ( \frac{ S_{34}}{\rm{KeVb}} \right ) \;
                   \left ( \frac{ S_{33}}{\rm{MeVb}} \right )^{-1/2}  \; T_6^{10}
\end{eqnarray}
where $T_6$ is the temperature in units of $10^6$K.

The high powers of $T$ in the above equations imply that the measured 
neutrino fluxes are strongly sensitive to $T$, i.e. $^7$Be 
and $^8$B neutrinos in 
principle are good thermometers for the  innermost part of the sun.

As we have seen in the previous section, ignoring  the possibility of sterile
neutrinos, by combining the results of SK and SNO one can determine
the total $^8$B neutrino flux, see eq.(\ref{flux}).
This can be used together with eq. (\ref{eq2ua}) to determine the central solar
temperature:
\begin{equation}
\label{eqT}
T = 15.7 \cdot(1\pm 1\%)\cdot 10^{6} \;{\rm K} 
\end{equation}
where the error gets comparable contributions from the uncertainty on $\Phi_{\rm B}$
and on nuclear physics.

This experimental resul is in excellent agreement with the prediction of recent
SSMs, see Table \ref{Tablessm}.

We add that, if sterile neutrinos are allowed, eq. (\ref{flux}) provides
a lower limit on the total neutrino flux, and correspondengly. eq. (\ref{eqT})
represents a lower limit to the central solar temperature.


\begin{table}
\caption[bb]{Predictions of some Solar Model calculations:}
\small
\begin{center}
\vspace{5 mm}
\begin{tabular}{||c|c|c|c|c|c||}
\hline \hline
&  BP2000 \cite{BP01} & FRANEC97\cite{FRANEC97} & RCVD96 \cite{RCVD96} 
& JCD96 \cite{sunjcd}& GARSOM2 \cite{Schlattl}\\ 
\hline
$T_c \; [10^6 \rm{K}]$ &  15.696   &  15.69 & 15.67 & 15.668 & 15.7\\  
$\Phi_{\rm B} \; [10^{6}{\rm cm}^{-2} {\rm s}^{-1}]$
                       &  5.05  &  5.16 &   6.33 & 5.87 &  5.30 \\
$\Phi_{\rm Be} \; [10^{9}{\rm cm}^{-2} {\rm s}^{-1}]$
                       &  4.77  &  4.49  &   4.8  & 4.94 &  4.93 \\
\hline \hline
\end{tabular}
\end{center}
\label{Tablessm}
\end{table}

Finally, one can use the above equations for getting an estimate of the 
$^7$Be neutrino flux in terms of experimental quantities.
From eqs. (\ref{eq2a}) and (\ref{eq2b}) one has:
\begin{equation}
\label{eqfibe}
\Phi_{\rm Be} = \frac{c_{Be}} {c_B^{1/2}}
              \left ( \Phi_{\rm B} \;  \frac{S_{34}} {S_{17} \sqrt{S_{33}}} \right )^{1/2} \quad .
\end{equation}
This gives:
\begin{equation}
\label{eqfibe2}
\Phi_{\rm Be}=4.9\,(1\pm 11\%) \cdot 10^9 \;{\rm cm}^{-2} {\rm s}^{-1}\quad
\end{equation}
where  again the uncertainty on the $^8$B neutrino flux and on nuclear physics
give comparable contribution to the error.

This estimate is as  accurate  \cite{BP01} as the SSM prediction 
and relies mainly on observational data.

\section{Acknowledgements}
It is a pleasure to thanks E. Lisi, G.L. Fogli and A. Palazzo for
a fruitful collaboration in recent years.

We are extremely grateful to M. Baldo Ceolin, N. Costa and G. Fogli
for organizing a very enjoyable, timely and interesting 
conference.


\begin{thebibliography}{99}

\bibitem{Hom}
B. Cleveland  et al., Ap. J. {\bf 496}, 505 (1998).

\bibitem{Gal}
GNO collaboration Phys. Lett. B {\bf 490} (2000) 16-26.

\bibitem{Sage}
SAGE collaboration,
Phys.Rev.Lett. 83 (1999) 4686-4689


\bibitem{Kam}
K.S. Hirata et al. (Kamiokande coll.),
 Phys. Rev. Lett. {\bf 77}, 1683 (1996).



\bibitem{SK}
S. Fukuda et al. (Super-Kamiokande coll.),
Phys. Rev. Lett. {\bf 86}, 5651 (2001).



\bibitem{Report} V. Castellani, S. Degl'Innocenti, G. Fiorentini and B.Ricci
Phys. Rep. {\bf 281}, 309, (1997).



\bibitem{SNO} G. Aardsma et al. (SNO coll.), Phys. Lett. B
 {\bf 194}, 321 (1987); Q.R. Ahmad et al. (SNO coll.) nucl-ex/0106015.


\bibitem{tio3}
F.L. Villante, G. Fiorentini, E. Lisi,
Phys. Rev. D {\bf 59}, 013006 (1999).

F.L. Villante, PhD thesis, Univesit\`a di Ferrara (2000).

\bibitem{elio} 
G.L. Fogli, E. Lisi, D. Montanino, A. Palazzo,
hep-ph/0106247

\bibitem{giunti} 
C. Giunti, astro-ph/0107310


\bibitem{venya}
V. Berezinsky, astro-ph/0108166

\bibitem{Fogli}
G.L. Fogli, E. Lisi, A. Palazzo, F.L. Villante,
Phys. Rev. D {\bf 63}, 113016 (1999).

\bibitem{BP01}
J.N. Bahcall, M. Pinsonneault and S. Basu,
Astrophys.J. {\bf 555} (2001) 990-1012



\bibitem{Bspe}
J.N. Bahcall, E.Lisi, D.E. Alburger, L. De Braeckeleer, S.J. Freedman and J. Napolitano,
Phys. Rev, C {\bf 54}, 411 (1996).

\bibitem{EScs}
J.N. Bahcall, M. Kamionkowsky and A. Sirlin,
Phys. Rev. D {\bf 51}, 6146 (1995).

\bibitem{CCcs}
K.\ Kubodera and S.\ Nozawa, 
Int.\ J.\ Mod.\ Phys.\ E {\bf 3}, 101 (1994).
We use the computer programs described in 
J.\ N.\ Bahcall and E.\ Lisi, 
Phys.\ Rev.\ D {\bf 54}, 5417 (1996)
and available at the URL
http://www.sns.ias.edu/$^\sim$jnb/SNdata~.

\bibitem{SKre}
Y. Suzuki, in {\em Neutrino~'98}, 
Proceedings of the XVIII International Conference on Neutrino Physics
and Astrophysics, Takayama, Japan.





\bibitem{Hi87}  S.\ Hiroi, H.\ Sakuma, T.\ Yanagida, and M.\ Yoshimura,
                Phys.\ Lett.\ B {\bf 198}, 403 (1987)
                        




\bibitem{BPBC}
J.N. Bahcall, M.H. Pinsonneault, S. Basu and J. Christensen-Dalsgaard,
Phys. Rev. Lett. {\bf 78}, 171 (1997).


\bibitem{Adelberger}
E. G. Adelberger et al., Rev. Mod. Phys. {\bf 70}, 1265 (1998)

\bibitem{Nacre}
C. Angulo et al. (Nacre coll.),
Nucl. Phys. A {\bf 656}, 3 (1999).


\bibitem{s17direct}
F. Hammache et al., Phys. Rev. Lett. {\bf 86} 3985 (2001)

\bibitem{s17indirect}
B. Davids et al., nucl-ex/0101010 (2001)

\bibitem{s33luna}
R. Bonetti et al. (LUNA coll.) Phys. Rev. Lett. {\bf 82} 5205 (1999);//
M.Junker et al. (LUNA coll.) Phys. Rev. C {\bf 57} 2700 (1998).




\bibitem{FRANEC97}
F. Ciacio, S. Degl'Innocenti and B. Ricci, Astron. Astroph. Suppl. {\bf 123}, 449 (1997).

\bibitem{RCVD96}
O. Richard, S. VAuclair, C. Charbonnel and W. A. Dziembowski
Astron. Astrph. {\bf 312} 1000 (1996)

\bibitem{sunjcd}
J. Christensen-Dalsgaard et al. , Science {\bf 272}  1286 (1996).

\bibitem{Schlattl}
H.Schlattl, A. Weiss and H.G. Ludwig, Astron. Astroph. {\bf 322}, 646 (1997).












%



\end{thebibliography}
\end{document}